\documentclass[11pt,a4paper]{article}
\pdfoutput=1

\usepackage[utf8]{inputenc}
\usepackage{a4wide}
\usepackage{amsmath}
\usepackage{cite}
\usepackage{xcolor}
\usepackage{amssymb}
\usepackage{amsfonts}

\usepackage{booktabs}
\usepackage{makecell}
\usepackage{multirow}
\usepackage{pdflscape}
\usepackage{textcomp}
\usepackage{gensymb}
\usepackage{bigstrut}
\usepackage{array}
\usepackage{enumerate}
\usepackage[small,bf]{caption}
\usepackage{subfig}
\usepackage{subcaption}
\usepackage{diagbox}
\usepackage{graphicx}
\usepackage[colorlinks=true,urlcolor=myred,anchorcolor=black,citecolor=myred,filecolor=black,linkcolor=myred,menucolor=blue,linktocpage=true]{hyperref}
\usepackage{appendix}
\usepackage{verbatim}
\usepackage{geometry}
\usepackage{listings}
\usepackage{mathtools}
\usepackage{longtable}
\usepackage{threeparttable}
\usepackage{floatpag}
\usepackage{framed}
\usepackage{calrsfs}
\DeclareMathAlphabet{\pazocal}{OMS}{zplm}{m}{n}
\SetMathAlphabet\pazocal{bold}{OMS}{zplm}{bx}{n}
\usepackage{multicol}

\usepackage[capitalise]{cleveref} 
\crefname{table}{Table}{Tables}
\crefname{figure}{Fig.}{Figs.}

\usepackage{xspace}

\usepackage{dcolumn}

\definecolor{myred}{rgb}{0.8, 0, 0.4}

\newcommand{\m}{{|\langle m \rangle|}}
\newcommand{\mno}{{|\langle m \rangle|_\text{NO}}}
\newcommand{\mio}{{|\langle m \rangle|_\text{IO}}}
\newcommand{\hl}{{T_{1/2}^{0\nu}}}
\newcommand{\bbnn}{{(\beta\beta)_{0\nu}}}

\newcommand{\at}{\alpha_{21}}
\newcommand{\att}{\alpha_{31}}
\newcommand{\atp}{\alpha_{31}'}

\newcommand{\ms}{\Delta m^2_\odot}
\newcommand{\ma}{\Delta m^2_\text{A}}

\newcommand{\mmin}{m_\text{min}}

\newcommand{\mref}{\m_0}

\newcommand{\ltap}{\ \raisebox{-.4ex}{\rlap{$\sim$}} \raisebox{.4ex}{$<$}\ }

\allowdisplaybreaks
\numberwithin{equation}{section}



\begin{document}

\renewcommand*{\thefootnote}{\fnsymbol{footnote}}
\begin{center}
{\bf\Large The meV frontier of neutrinoless
double beta decay\\[2mm] in the JUNO era
}\\[8mm]
J.~T.~Penedo$^{\,a~}$\footnote{E-mail: \href{jpenedo@roma3.infn.it}{\texttt{jpenedo@roma3.infn.it}}},
S.~T.~Petcov$^{\,b,c~}$\footnote{E-mail: \href{petcov@sissa.it}{\texttt{petcov@sissa.it}}. Also at Institute of Nuclear Research and Nuclear Energy, Bulgarian Academy of Sciences, 1784 Sofia, Bulgaria.}\\
 \vspace{5mm}
$^{a}$\,{\it INFN Sezione di Roma Tre, Via della Vasca Navale 84, 00146, Roma, Italy} \\[1mm]
$^{b}$\,{\it \small SISSA/INFN, Via Bonomea 265, 34136 Trieste, Italy} \\[1mm]
$^{c}$\,{\it \small Kavli IPMU (WPI), UTIAS, The University of Tokyo, Kashiwa, Chiba 277-8583, Japan}\\
\vspace{2mm}

\end{center}
\vspace{2mm}

\begin{abstract}
Observing neutrinoless double beta decay would
establish lepton number violation and the Majorana nature of neutrinos.
Within the standard 3-flavour paradigm, the rate of this process is controlled by the
effective Majorana mass $\m$, which may be severely suppressed if the 
neutrino mass spectrum presents normal ordering.
Taking into account the first JUNO results, which significantly reduce the uncertainties on solar neutrino oscillation parameters, 
we provide updated conditions
under which $\mno$ is guaranteed to  
exceed the $10^{-3}$ eV ($5\times 10^{-3}$ eV) threshold.
We analyse both the generic case, as well as scenarios where
the two Majorana phases either take CP conserving values, or
at least one of them takes a CP-violating value,
that are in line with predictive 
schemes combining flavour 
and generalised CP symmetries.
\end{abstract}

\renewcommand*{\thefootnote}{\arabic{footnote}}
\setcounter{footnote}{0}

%
\section{Introduction}


Neutrino physics has entered an era of consolidation and precision. A wide range of solar, atmospheric, reactor, and accelerator neutrino data are consistently described within the standard three-neutrino paradigm~\cite{ParticleDataGroup:2024cfk,Capozzi:2025wyn}.
However, it is known that neutrino oscillations are insensitive
to the absolute neutrino mass scale and to the nature of massive neutrinos (Dirac vs.~Majorana)~\cite{Bilenky:1980cx,Langacker:1986jv}. A crucial probe of these properties is neutrinoless double beta \mbox{($\bbnn$-)decay}: a lepton-number-violating transition between isobars $(A,Z)$ and $(A,Z+2)$, accompanied by the emission of two electrons, but notably lacking the emission of the two electron anti-neutrinos (see, e.g.~the reviews~\cite{Bilenky:1987ty,Gomez-Cadenas:2023vca}).
The \mbox{$\bbnn$-decay} rate is sensitive also to the type of 
neutrino mass spectrum -- with normal ordering (NO) or inverted 
ordering (IO) -- that neutrino masses obey~\cite{Pascoli:2002xq}.

Currently, the best limits on the $\bbnn$-decay half-lives have been obtained for the isotopes of germanium-76, tellurium-130, and xenon-136:
\begin{itemize}
    \item $\hl(^{130}\text{Te}) > 3.5 \times 10^{25}\text{ yr}$, by the CUORE collaboration~\cite{CUORE:2024ikf},  
    \item $\hl(^{76}\text{Ge}) > 1.8 \times 10^{26}\text{ yr}$, by the GERDA collaboration (phases I+II)~\cite{GERDA:2020xhi}, and
    \item $\hl(^{136}\text{Xe}) > 3.8 \times 10^{26}\text{ yr}$, by the KamLAND-Zen collaboration~\cite{KamLAND-Zen:2024eml},%
    \footnote{
    The weaker bound $\hl(^{136}\text{Xe}) > 3.5 \times 10^{25}\text{ yr}$ (90\% CL) has been independently obtained by the EXO-200 collaboration~\cite{EXO-200:2019rkq}.
    }
\end{itemize}
where all bounds are given at the 90\% confidence level (CL).

In this paper we revisit the meV frontier of \mbox{$\bbnn$-decay},
discussed in~\cite{Penedo:2018kpc,Pascoli:2007qh},
taking into account the first JUNO data release~\cite{JUNO:2025gmd} and
its impact on the most recent neutrino global
fits~\cite{Capozzi:2025wyn,Capozzi:2025ovi}.
We focus on the standard scenario, where
the exchange of three light Majorana neutrinos 
$\nu_i$ $(i=1,2,3)$
provides the dominant contribution to the decay rate,
which is proportional to the 
square of the
so-called effective Majorana mass $\m$.
While $\m$ is bounded from below for a neutrino mass
spectrum with inverted ordering (IO),
it can instead be vanishingly small
in the case of a spectrum with normal ordering (NO).
Here, we spell out the updated conditions under which the 
effective Majorana mass exceeds the 
$10^{-3}$ eV ($5\times 10^{-3}$ eV) threshold, 
which is the goal of beyond-the-ton-scale 
future \mbox{$\bbnn$-decay} experiments
(see, e.g.,~\cite{guenette_2024_12706038}).
Beyond the generic case, corresponding to unconstrained Majorana
CP violation (CPV) phases, we analyse
a set of cases in which the CPV phases take values motivated by predictive symmetric schemes, namely those combining generalised CP (gCP) and flavour symmetries.

\vskip 2mm
An extensive analysis of the implications of the first JUNO results
for the phenomenology of the effective Majorana mass was recently presented
in Ref.~\cite{Ge:2025cky}, where the first JUNO data are used to refine the
allowed parameter space and probability distributions for $\m$.
In contrast, and in continuity with Ref.~\cite{Penedo:2018kpc},
the present study focuses on revisiting the general conditions
under which $\m$ exceeds the aforementioned thresholds,
either independently of or for fixed values of the Majorana phases.
While some numerical results, such as the minimal values of $\m$ in the NO and IO cases, naturally overlap and are consistent across the two analyses, the focus and scope of this work remain distinct from those of Ref.~\cite{Ge:2025cky}.

%
\section{The effective Majorana mass}
%

In the standard three light neutrino scenario, the inverse of the decay
half-life reads:
\begin{equation}
\big(T_{1/2}^{0\nu}\big)^{-1} \,= \,G_{0\nu}(Q,Z) \,
\big|M_{0\nu}(A,Z)\big|^2\, \m^2 \,,
\end{equation}
%
where $G_{0\nu}$ denotes the phase-space factor,
$M_{0\nu}$ is the
nuclear matrix element (NME) of the decay,%
\footnote{NMEs make up the main source of uncertainty in extracting
$\m$ from data. See e.g.~\cite{Miramonti:2025ges}
for a recent summary of the progress in experimental techniques and
NME calculations.
}
and $\m$ is the effective Majorana mass, given by
(see, e.g.~\cite{Bilenky:1987ty}):
\begin{equation}
  \label{eq:meff}
\textstyle
\m \,=\, \left|\,\sum_{i=1}^3 U_{\text{e}i}^2 \,m_i\,\right|\,.
\end{equation}
%
Here, 
$m_i$ are the neutrino masses and
$U_{\text{e}i}$ are elements of the first row of the
Pontecorvo-Maki-Nakagawa-Sakata (PMNS) leptonic mixing matrix.
In the standard parameterization~\cite{ParticleDataGroup:2024cfk},
the latter read
\begin{equation}
  \label{eq:Uei}
U_{\text{e}i} = 
\begin{pmatrix}
c_{12}\, c_{13}\,, &
s_{12}\, c_{13}\, e^{i\alpha_{21}/2}\,, &
s_{13}\, e^{i\atp/2\,}
\end{pmatrix}_i \,,
\end{equation}
%
with $c_{ij} \equiv \cos \theta_{ij}$ and $s_{ij} \equiv \sin \theta_{ij}$,
where $\theta_{ij} \in [0,\pi/2]$ are the lepton mixing angles.
We have further defined an effective Majorana CPV phase
$\atp \equiv \att - 2\delta$, with
$\delta$ and the $\alpha_{ij}$ 
being the standard Dirac and Majorana~\cite{Bilenky:1980cx} CPV phases,
respectively ($\delta, \alpha_{ij} \in [0,2\pi)$).

The most stringent upper limit on $\m$
was obtained by the KamLAND-Zen collaboration,
using the previously quoted lower limit on
$\hl(^{136}\text{Xe})$~\cite{KamLAND-Zen:2024eml}:
\begin{equation}
\m < (0.028-0.122)~{\rm eV}\,,~~90\%~\text{CL}\,,
\label{eq:meffKZ}
\end{equation}
%
which considers a range of phenomenological NME calculations.
Taking also into account 
the effects of short-range contact contributions~\cite{Cirigliano:2018hja}
(see also~\cite{Jokiniemi:2021qqv,Weiss:2021rig,Jokiniemi:2022ayc,Belley:2022zsq})
leads to updated NME determinations~\cite{Jokiniemi:2022ayc,Belley:2023btr}
and to the wider range~\cite{KamLAND-Zen:2024eml}:
\begin{equation}
\m < (0.025-0.138)~{\rm eV}\,,~~90\%~\text{CL}\,.
\label{eq:meffKZ-sr}
\end{equation}
%
This range is obtained from the union of
ab initio, quasiparticle random phase approximation, and shell model
results.

\vskip 2mm

At present, neutrino oscillation data provides precise information on the leptonic mixing angles $\theta_{ij}$ and on the solar ($\ms$) and atmospheric ($\ma$) neutrino mass-squared differences.
From~\cref{eq:meff,eq:Uei}, it is clear that the rate of 
\mbox{$\bbnn$-decay} crucially depends on four unknowns: the lightest neutrino mass $\mmin$, the two (effective) Majorana CPV phases $\alpha_{ij}^{(\prime)}$, and the type of neutrino mass spectrum (NO vs.~IO).%
\footnote{An NO or IO mass spectrum is said to be normal 
hierarchical (NH) or inverted hierarchical (IH) if
$m_1 \ll m_{2,3}$ or $m_3 \ll m_{1,2}$, respectively.
The opposite limit of relatively large $\mmin$ corresponds to a 
quasi-degenerate (QD) spectrum, with $m_1 \simeq m_2 \simeq m_3$.}
At present, global fits of neutrino data show a mild preference for NO over IO, at the $2.2\sigma$ level in the analysis of Ref.~\cite{Capozzi:2025wyn}.
In the prevalent convention, one has $\ms \equiv \Delta m^2_{21} > 0$ and:
\begin{itemize}
\item $m_1 < m_2 < m_3$, $\Delta m^2_{31} \equiv \ma>0$, for NO,
\item $m_3 < m_1 < m_2$, $-\Delta m^2_{23} \equiv \ma<0$, for IO,
\end{itemize}
where $\Delta m^2_{ij} \equiv m^2_i- m^2_j$.
For either ordering, $|\ma| = \max (|m_i^2-m_j^2|)$, with $i,j=1,2,3$.
One has $\mmin \equiv m_1\, (m_3)$ in the NO (IO) case.
The effective Majorana mass can then be explicitly written as:
\begin{align} \label{eq:meffNO}
\mno =&\,
\bigg|\mmin\,c_{12}^2\,c_{13}^2+\sqrt{\ms + \mmin^2}\, s_{12}^2\,c_{13}^2
\,e^{i\alpha_{21}} + \sqrt{\ma + \mmin^2}\, s_{13}^2\,e^{i\alpha_{31}'}\bigg|
\,,\\ \label{eq:meffIO}
\mio =&\, 
\bigg|\sqrt{|\ma|-\ms+\mmin^2}\,c_{12}^2\,c_{13}^2 
+ \sqrt{|\ma| + \mmin^2}\, s_{12}^2\,c_{13}^2 \,e^{i\alpha_{21}} 
+ \mmin\, s_{13}^2\,e^{i\alpha_{31}'}\bigg|
\,,
\end{align}
depending on the ordering.
From these expressions, one sees that the effective Majorana mass can be interpreted as the length
of the sum of three vectors in the complex plane,
with relative orientations given by $\at$ and $\atp$. See, e.g.~\cite{Penedo:2018kpc} for further details on this geometric interpretation.

\vskip 2mm

The Jiangmen Underground Neutrino Observatory (JUNO) reactor neutrino experiment~\cite{JUNO:2015zny,JUNO:2022mxj,JUNO:2024jaw},
conceived in~\cite{Petcov:2001sy,Choubey:2003qx,Learned:2006wy,Li:2013zyd},
has recently started to take data. JUNO is designed to determine the neutrino mass ordering and is expected to measure the solar angle $\theta_{12}$ as well as $\ms$ and $|\ma|$ with unprecedented precision~\cite{Choubey:2003qx}. Using only the first 59.1 days of data~\cite{JUNO:2025gmd}, JUNO has managed to reduce the uncertainty on solar parameters by a factor of $\sim 1.5$ with respect to the global fit of Ref.~\cite{Capozzi:2025wyn}, see~\cite{Capozzi:2025ovi}.
In what follows, we revisit
the results of our 2018 work~\cite{Penedo:2018kpc} taking into account
the latest global fits~\cite{Capozzi:2025wyn,Capozzi:2025ovi},
incorporating JUNO data.%
\footnote{See also the independent results obtained by the NuFIT group~\cite{Esteban:2024eli}, and the corresponding update incorporating JUNO data~\cite{Esteban:2026phq,nufit-6.1}.}
The relevant data, namely
the $n\sigma$ ranges for 
$\Delta m^2_{32}$, $\Delta m^2_{21}$,
$\sin^2 \theta_{12}$, and $\sin^2 \theta_{13}$,
with $n=3$ for IO and $n=1,2,3$ for NO,
are summarised in~\cref{tab:dataIO,tab:dataNO}, respectively.

\begin{table}[t]
\centering
\begin{tabular}{cccc}
\toprule
$\dfrac{\Delta m^2_{21}}{10^{-5} \text{ eV}^2}$ & 
$\dfrac{\Delta m^2_{23}}{10^{-3} \text{ eV}^2}$ &
$\dfrac{\sin^2 \theta_{12}}{10^{-1}}$ &
$\dfrac{\sin^2 \theta_{13}}{10^{-2}}$ \\[5pt]
\midrule
\rule{0pt}{10pt}
$7.21-7.78$ & $2.44-2.56$ & $2.87-3.30$ & $2.08-2.41$
\\ \bottomrule
\end{tabular}
\caption{\label{tab:dataIO}%
Ranges for the oscillation parameters of interest in the IO case, at the $3\sigma$ CL, 
from Refs.~\cite{Capozzi:2025wyn,Capozzi:2025ovi}.
The atmospheric mass-squared difference $\Delta m^2_{23}$ is obtained from the quantities defined in~\cite{Capozzi:2025wyn} using the best-fit value of $\Delta m^2_{21}$,
namely $\Delta m^2_{32} = \Delta m^2 - \delta m^2 / 2$,
where $\delta m^2 = \Delta m^2_{21}$.
}
\end{table}
\begin{table}[t]
\centering
\vskip 2mm
\begin{tabular}{lccc}
\toprule
Parameter&
$1\sigma$ range&
$2\sigma$ range&
$3\sigma$ range\\
\midrule
\rule{0pt}{10pt}%
$\Delta m^2_{21}\,/\,(10^{-5} \text{ eV}^2)$ & $7.39-7.58$ & $7.30-7.68$ & $7.21-7.78$ \\
$\Delta m^2_{31}\,/\,(10^{-3} \text{ eV}^2)$ & $2.51-2.55$ & $2.49-2.57$ & $2.47-2.60$ \\
$\sin^2 \theta_{12}\,/\,10^{-1}$ & $3.01-3.16$ & $2.94-3.23$ & $2.87-3.30$ \\
$\sin^2 \theta_{13}\,/\,10^{-2}$ & $2.17-2.27$ & $2.11-2.33$ & $2.06-2.38$
\\ \bottomrule
\end{tabular}
\caption{\label{tab:dataNO}%
Ranges for the oscillation parameters of interest in the NO case, at the  $n\sigma$ ($n=1,2,3$) CLs,
from Refs.~\cite{Capozzi:2025wyn,Capozzi:2025ovi}.
The atmospheric mass-squared difference $\Delta m^2_{31}$ is obtained from the quantities defined in~\cite{Capozzi:2025wyn} using the best-fit value of $\Delta m^2_{21}$,
namely $\Delta m^2_{31} = \Delta m^2 + \delta m^2 / 2$,
with $\delta m^2 = \Delta m^2_{21}$.
}
\end{table}

We note that, from the above discussion,
it is clear that the upper bounds on $\m$
already imply a bound on $\mmin$ in the quasi-degenerate (QD) regime. One finds:
\begin{equation}
    \mmin \,\ltap\, \frac{0.122\text{ eV}\,\, (0.138\text{ eV})}{\cos 2\theta_{12} - \sin^2\theta_{13}}
    \,\ltap\, 0.39 \text{ eV}\,\, (0.44 \text{ eV})\,,
\end{equation}
%
obtained using the maximal KamLAND-Zen bound of~\cref{eq:meffKZ}
(of~\cref{eq:meffKZ-sr}, which
accounts for the short-range contribution in $\bbnn$-decay),
as well as the minimal and maximal
$3\sigma$ values of $\cos 2\theta_{12}$ and $\sin^2\theta_{13}$, respectively.
This bound on the lightest neutrino mass should be compared with other constraints on the absolute neutrino mass scale. In particular, these refer to i) tritium beta decay end-point measurements, which constrain $m_\beta^2 \equiv \sum_i |U_{\text{e}i}|^2\,m_i^2$, and ii) cosmological data, which inform instead on the sum of neutrino masses $\Sigma \equiv \sum_i m_i$. 
The most stringent bound on $m_\beta$
comes from the KATRIN experiment~\cite{Eitel:2005hg}
and reads 
$m_\beta < 0.45$ eV (90\% CL)~\cite{KATRIN:2024cdt}.
By varying the relevant
mixing angles and mass-squared differences within their $3\sigma$ ranges,
one can translate this bound into the constraint 
$\mmin < 0.45$ eV, for both orderings.
Regarding the constraint from neutrino cosmology,
note that
upper limits on $\Sigma$ vary significantly 
and depend on the choice of likelihood function,
dataset, and cosmological model.
In the absence of a consensus, we consider
the central, moderate bound
$\Sigma < 0.2$ eV quoted in Ref.~\cite{Capozzi:2025wyn},
which translates to $\mmin < 0.06\text{ eV}\,(0.05\text{ eV})$
in the NO (IO) case,
for $3\sigma$ variations of mass-squared differences.

%
\section{Unconstrained Majorana phases}
%
\subsection{The case of inverted ordering}

In the IO case,
the lengths of the three vectors entering~\cref{eq:meffIO} are mildly hierarchical and decreasing.
Therefore, extremal values of $\mio$ are obtained
when the three vectors are aligned ($\at=\atp=0$, maximizing $\m$)
or when the first vector is anti-aligned with the others
($\at=\atp=\pi$, minimizing $\m$).
It follows that there is a lower bound on $\mio$
for every value of $\mmin$~\cite{Pascoli:2002xq}.
Using the data in~\cref{tab:dataIO}, we find
$\mio > 1.58 \times 10^{-2}$ eV,
for variations of individual oscillation parameters
in their $3\sigma$ ranges.
In the limit of a negligible lightest neutrino mass, $\mmin\to 0$ (IH spectrum),
we find $\mio \in [1.6,4.9]\times 10^{-2}$ eV.

\subsection{The case of normal ordering}

In what follows, we concentrate on the NO case. 
Considering variations of oscillation parameters in their
$3\sigma$ ranges (see~\cref{tab:dataNO}),
for $\mmin < 5\times 10^{-2}$ eV ($< 1\times 10^{-2}$ eV) there is an \emph{upper} bound on
$\mno < 5.1\times 10^{-2}$ eV ($< 1.2 \times 10^{-2}$ eV), obtained for $\at=\atp=0$.
In the limit of negligible $\mmin$,
we find $\mno \in [1.2,4.1]\times 10^{-3}$ eV.

\vskip 2mm

Unlike in the IO case, the lengths of the complex vectors now allow for cancellations in $\mno$. Instead of a lower bound, and depending on the precise values of oscillation parameters, and especially on the values of CPV phases and $\mmin$, one may instead fall into a ``well of unobservability''.
\begin{figure}[t]
\centering
\subfloat[$\mref = 1$ meV]{\includegraphics[width=0.65\textwidth]{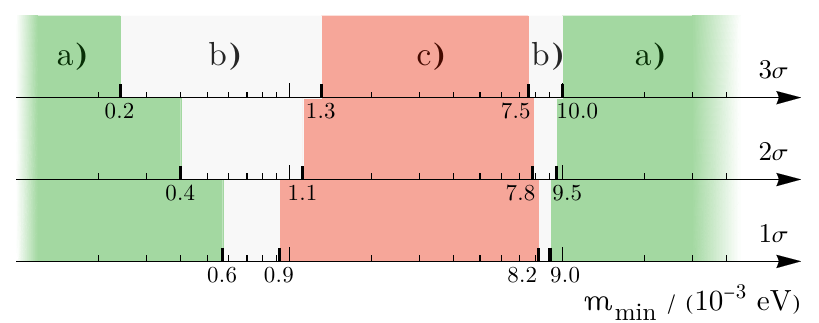}
\label{fig:th}}
\vskip 4mm
\subfloat[$\mref = 5$ meV]{\includegraphics[width=0.65\textwidth]{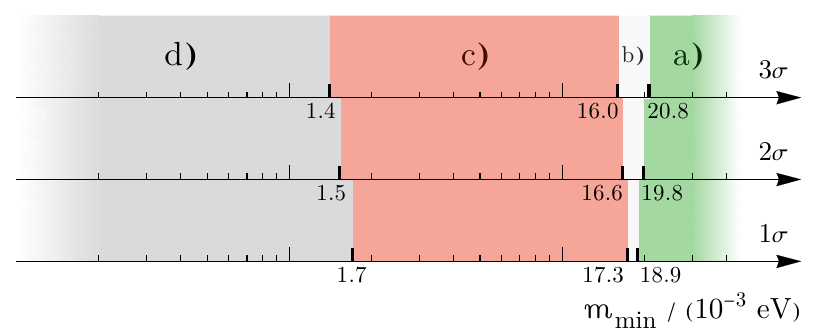}
\label{fig:th5}}
\caption{%
Ranges of the lightest mass $\mmin= m_1$ (NO spectrum)
for which different conditions apply: 
in green, a) $\mno > \mref$, for all values of parameters and phases; 
in light grey, b) there exist values of parameters and phases for which $\mno < \mref$;
in red, c) for all values of parameters, there exist phases such that
$\mno < \mref$;
in dark grey, d) $\mno < \mref$, for all values of parameters and phases.
See text for further details.
Oscillation parameters are varied within their $n\sigma$ ($n=1,2,3$) ranges, cf.~\cref{tab:dataNO}.
}
\label{fig:th0}
\end{figure}

Consider first the meV reference threshold, $\mref = 10^{-3}$ eV. 
By varying oscillation parameters in their respective $n\sigma$ ($n=1,2,3$) ranges and CPV phases in their full $[0,2\pi]$ intervals, different values of $\mmin$ -- corresponding to different neutrino absolute mass scales -- lead to different scenarios for $\mno$. One can have that: 
\begin{itemize}
    \item[a)] $\mno > \mref$ for all values of  
$\theta_{ij}$, $\Delta m^2_{ij}$, and $\alpha_{ij}^{(\prime)}$ 
within their allowed intervals;
    \item[b)] there exist specific values of $\theta_{ij}$, $\Delta m^2_{ij}$ 
within their $n\sigma$ intervals and
values of $\alpha_{21}$ and $\alpha_{31}'$ such that $\mno < \mref$;
    \item[c)] for all values of $\theta_{ij}$ and $\Delta m^2_{ij}$ 
from the corresponding $n\sigma$ intervals
there exist values of the phases $\alpha_{21}$ and $\alpha_{31}'$
for which $\mno < \mref$;
    \item[d)] $\mno < \mref$ for any values of oscillation parameters and CPV phases within their allowed intervals.
\end{itemize}
The results of this analysis are summarised in~\cref{fig:th}, where the
$\mmin$ intervals in which each condition applies are indicated, for different $n\sigma$ CLs.
As shown, for $3\sigma$ variations
of the $\sin^2 \theta_{ij}$ and $\Delta m^2_{ij}$,
one is guaranteed to have $\mno > 10^{-3}$ eV, provided either
$\mmin < 0.2 \times 10^{-3}$ eV
or $\mmin > 10.0 \times 10^{-3}$ eV.
These bounds on $\mmin$ translate, respectively,
into the conditions
$\Sigma < 0.060$ eV and $\Sigma > 0.074$ eV
on the sum of neutrino masses.
Conversely, if $\mmin \in [1.3,7.5] \times 10^{-3}$ eV,
there is \emph{always} a choice of $\at$
and $\atp$ such that $\mno < 10^{-3}$ eV, for oscillation parameters within their $3\sigma$ ranges (condition c)).
There is no value of $\mmin$ satisfying condition d) for this reference threshold. 

\vskip 2mm
The above analysis can be repeated for the higher reference value $\mref = 5\times 10^{-3}$ eV. The corresponding results are summarised in~\cref{fig:th5}.
As shown, for $3\sigma$ variations
of oscillation parameters,
$\mno > 5\times 10^{-3}$ eV is guaranteed provided $\mmin > 20.8 \times 10^{-3}$ eV.
This condition on $\mmin$ corresponds to having
a large enough sum of neutrino masses, $\Sigma > 0.097$ eV.
Instead, if $\mmin < 16.0 \times 10^{-3}$ eV, there is always a choice of $\at$
and $\atp$ such that $\mno < 5 \times 10^{-3}$ eV (for oscillation parameters within their $3\sigma$ ranges).

\begin{figure}[t]
\centering
\hspace{-2cm}
\includegraphics[width=8cm]{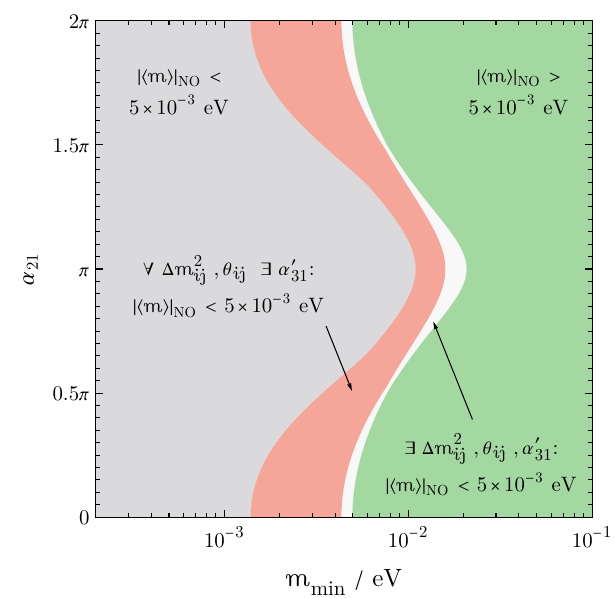}
\includegraphics[width=8cm]{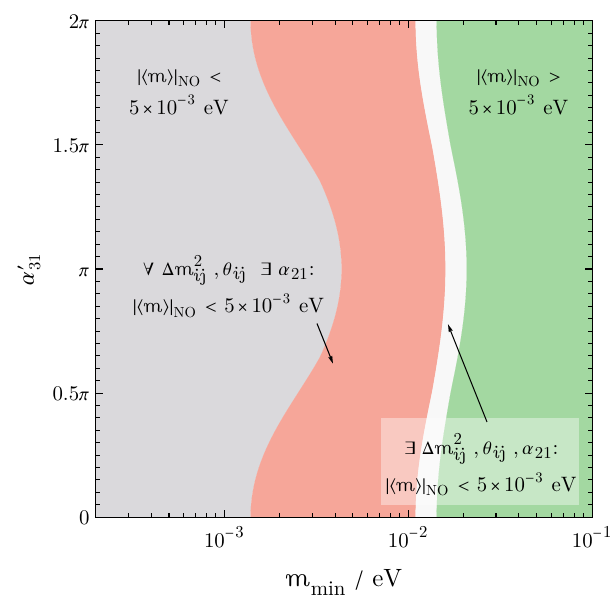}
\hspace{-1.3cm}
\caption{%
Regions in the $\mmin$--\,$\at$ plane (left)
and in the $\mmin$--\,$\atp$ plane (right)
where the different conditions on $\mno$ apply, for a reference value $\mref = 5\times 10^{-3}$ eV and $3\sigma$ variations of oscillation parameters.
Colours match those in~\cref{fig:th0}.
}
\label{fig:plane}
\end{figure}

To elucidate the role of the CPV phases, we additionally show in~\cref{fig:plane}
the regions in the $\mmin$--\,$\at$ and $\mmin$--\,$\atp$ planes
where the different conditions on $\mno$ apply
(now varying only the remaining phase),
for a reference value $\mref = 5\times 10^{-3}$ eV
and $3\sigma$ variations of oscillation parameters.
While the top row of~\cref{fig:th5} leaves it implicit, the dependence on
the phases is now made explicit. These plots show that the range of masses that allow to be above the threshold is extended for values of
$\at$ close to $0$ or $2\pi$,
while the effect of varying $\atp$ is comparatively milder.

%
\section{CP and generalised CP}
%

Since $\m$ strongly depends on  $\at$ and $\atp$, we now turn to scenarios in which these phases are strongly constrained.
Requiring CP invariance, for instance, directly constrains the leptonic CPV phases
to be integer multiples of $\pi$~\cite{Wolfenstein:1981rk,Kayser:1984ge,Bilenky:1984fg},
leading to $\at,\,\atp \in \{ 0,\,\pi\}$ (\emph{CP-conserving} values). 
Different but precise values for these phases can be obtained in frameworks where a discrete flavour symmetry is consistently combined with a generalised CP (gCP) symmetry~\cite{Feruglio:2012cw,Holthausen:2012dk},
corresponding to the most general form of CP~\cite{Branco:1986gr,Ecker:1987qp,Neufeld:1987wa,Grimus:1995zi}.
Ref.~\cite{Penedo:2018kpc} surveyed predictive schemes based on symmetry groups with $<100$ elements, singling out the values
$\at,\,\atp \in \{0,\, \pi/2,\, \pi,\,3\pi/2\}$ as a more general set of sharp predictions to consider (\emph{gCP-compatible} values). Clearly, some but not all of these gCP-compatible values are CP conserving in spite of the imposed gCP symmetry~\cite{Feruglio:2012cw}.

\begin{figure}[t]
\centering
\includegraphics[width=14cm]{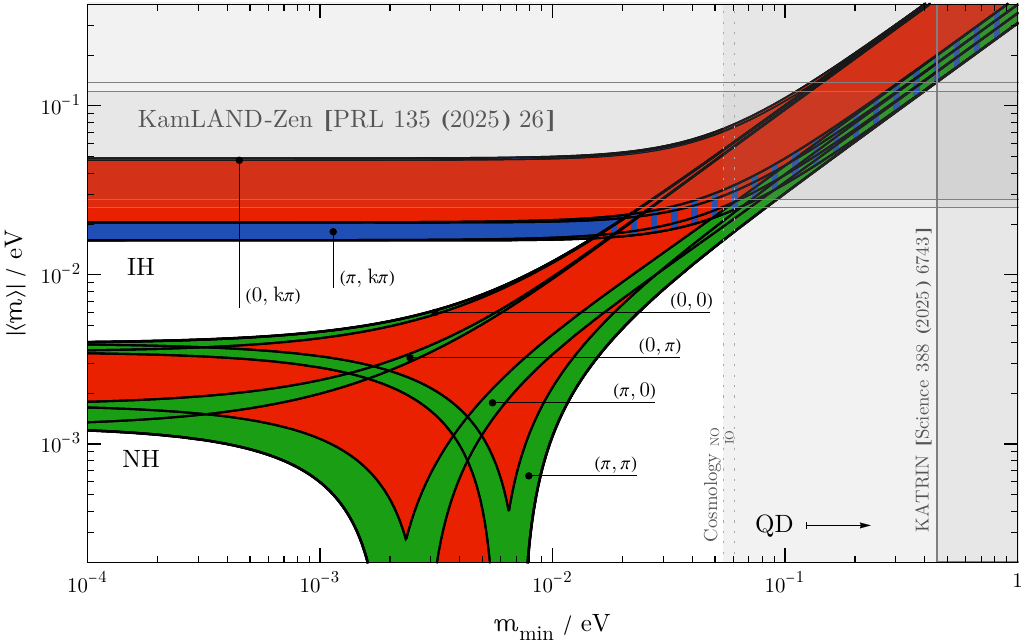}
\caption{%
The effective Majorana mass $\m$ as a function of $\mmin$, for both orderings and all possible values of the CPV phases $\at$ and $\atp=\att-2\delta$,
at the $2\sigma$ CL.
Green (blue) bands correspond to
CP-conserving values of the phases, for a spectrum with NO (IO), with $k=0,1$.
Within the red regions, at least one of the phases takes a CP-violating value.
The KamLAND-Zen bounds of~\cref{eq:meffKZ,eq:meffKZ-sr}, from~\cite{KamLAND-Zen:2024eml}, as well as the discussed KATRIN~\cite{KATRIN:2024cdt} and cosmology~\cite{Capozzi:2025wyn} constraints on $\mmin$, are shown.
}
\label{fig:CP}
\end{figure}

\vskip 2mm

We now look into the behaviour of $\mno$ and $\mio$ for each
of the 16 possible pairs $(\at,\,\atp)$ of gCP-compatible phases.
Some of these are equivalent, since they are related by conjugation,
leading to the same value of $\m$. One is left with
10 inequivalent pairs of interest,
\begin{equation}
\begin{aligned}
(\at,\,\atp) \in
\{
&(0,\,0),\,
(0,\,\pi/2),\,
(0,\,\pi),\,
(\pi/2,\,0),\,
(\pi/2,\,\pi/2),\,\\
&(\pi/2,\,\pi),\,
(\pi,\,0),\,
(\pi,\,\pi/2),\,
(\pi,\,\pi),\,
(3\pi/2,\,\pi/2)\}\,.
\end{aligned}    
\end{equation}
In~\cref{fig:CP}, we show the $2\sigma$-allowed regions for the effective Majorana mass $\m$, for both orderings. 
These regions are found using an approximate $\chi^2$ function, built from the sum of the one-dimensional projections of Refs.~\cite{Capozzi:2025wyn,Capozzi:2025ovi}.
The subregions corresponding to CP-conserving pairs of phases are highlighted in green (NO) and blue (IO) and delimit the allowed parameter space.

\begin{figure}[t]
\centering
\includegraphics[width=9.5cm]{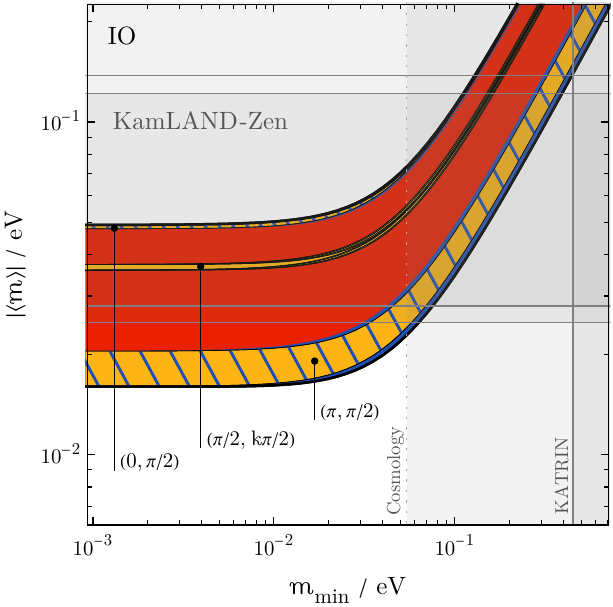}
\caption{%
The effective Majorana mass $\m$ as a function of $\mmin$, for a spectrum with IO and all possible values of the CPV phases,
at the $2\sigma$ CL.
Blue bands correspond to CP-conserving values of phases (cf.~\cref{fig:CP}),
while yellow bands refer to pairs of phases that are gCP-compatible
but not fully CP-conserving ($k=0,1,2,3$). Hatching indicates the overlap of blue and yellow regions, occurring for the cases $(0,\pi/2)$ and $(\pi,\pi/2)$.}
\label{fig:IO_gCP}
\end{figure}

\begin{figure}[p]
\centering
\includegraphics[width=13cm]{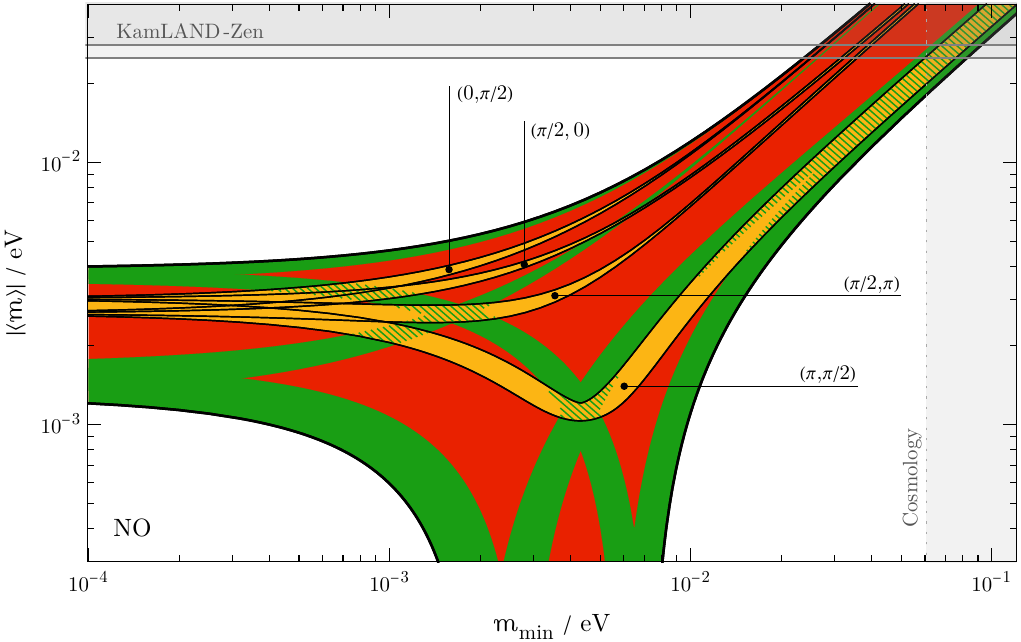}
\caption{%
The effective Majorana mass $\m$ as a function of $\mmin$, for a spectrum with NO and all possible values of the CPV phases,
at the $2\sigma$ CL.
Green bands correspond to CP-conserving values of phases (cf.~\cref{fig:CP}),
while yellow bands refer to pairs of gCP-compatible phases, with exactly one of them being CP-conserving. Hatching indicates region overlap.
}
\label{fig:NO_gCP1}
\end{figure}

\begin{figure}[p]
\centering
\includegraphics[width=13cm]{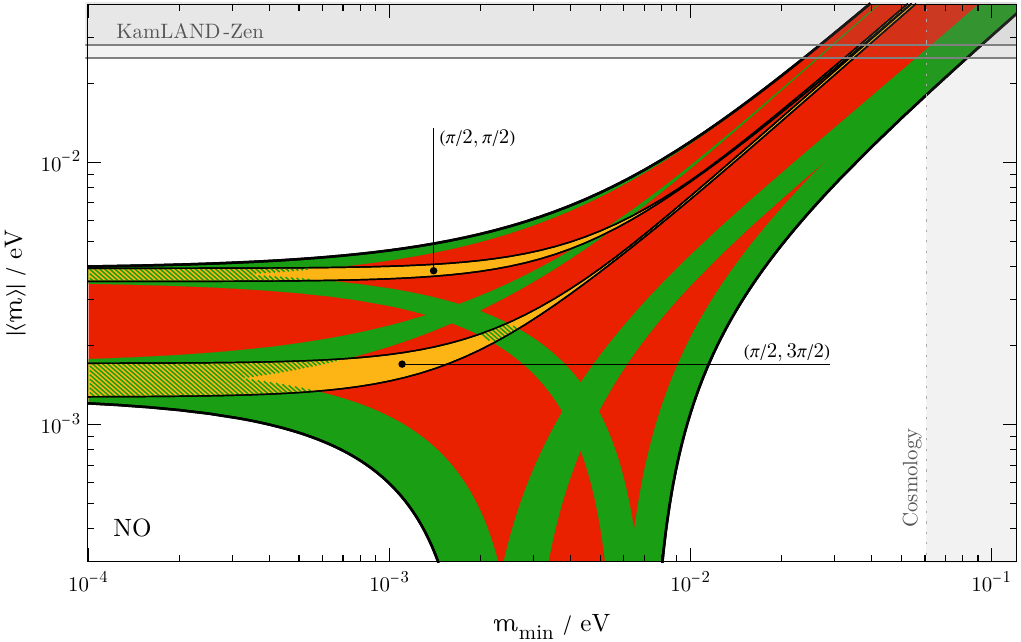}
\caption{%
The same as~\cref{fig:NO_gCP1}, with 
yellow bands referring to pairs of gCP-compatible phases where none of them are CP-conserving.}
\label{fig:NO_gCP2}
\end{figure}

Cases where at least one phase is gCP-compatible but not CP-conserving are presented in~\cref{fig:IO_gCP} for IO,%
\footnote{Note that, in the case of IO, there is substantial overlap between
the $(0,\,0)$ and $(0,\,\pi)$ bands and
between the $(\pi,\,0)$ and $(\pi,\,\pi)$ bands,
as seen in~\cref{fig:CP},
as well as between the four $(\pi/2,\,k\,\pi/2)$ yellow bands with $k=0,1,2,3$, in~\cref{fig:IO_gCP}.}
and in~\cref{fig:NO_gCP1,fig:NO_gCP2} for NO.
These results show that there exist values of the effective Majorana mass
which are incompatible with CP conservation, but may still be compatible with a gCP-based scheme.
For the NO case, we also summarise in~\cref{tab:gCP} the lower bounds on $\mno$, for each pair of phases. It is interesting to note that
$\mno$ is bounded from below at the $3\sigma$ CL, with $\mno \geq 1$ meV, in all the cases where the Majorana phases are gCP-compatible but not CP-conserving.

\begin{table}[t]
\centering
\begin{threeparttable}
\begin{tabular}{lcccc}
\toprule
& \multicolumn{4}{c}{$\atp$}\\
\cline{2-5}
\rule{0pt}{10pt}%
$\at\qquad$ & $0$ & $\pi/2$ & $\pi$ & $3\pi/2$ \\
\midrule
\rule{0pt}{10pt}%
$0$      & $3.4\,(3.5)$        & $2.6\,(2.7)$    & $1.2\,(1.3)$          & $\sim(0,\,\pi/2)$ \\
$\pi/2$  & $2.6\,(2.7)$        & $3.4\,(3.5)$    & $2.4\,(2.4)$
         & $\sim(3\pi/2,\,\pi/2)$ \\
$\pi$    & no bound\tnote{a}
         & $1.0\,(1.0)$
         & no bound\tnote{b}
         & $\sim(\pi,\,\pi/2)$ \\
$3\pi/2$ & $\sim(\pi/2,\,0)$ & $1.2\,(1.3)$    & $\sim(\pi/2,\,\pi)$ & $\sim(\pi/2,\,\pi/2)$ 
\\ \bottomrule
\end{tabular}
    \begin{tablenotes}
      \item[a]{$\mno > 1$ meV if $\mmin \notin [0.3,4.9]\, ([0.4,4.7])$ meV.}
      \item[b]{$\mno > 1$ meV if $\mmin \notin [3.8,10.1]\,([4.0,9.7])$ meV.}
    \end{tablenotes}
\end{threeparttable}
\caption{\label{tab:gCP}%
Lower bounds on $\mno$  in meV, at the $3\sigma$ ($2\sigma$) CL, for different fixed values of the phases $\at$ and $\atp$. A tilde denotes equivalence between cases.
For cases without a lower bound, the condition on $\mmin$ leading to $\mno>1$ meV is indicated.
}
\end{table}

%
\section{Conclusions}
%
%
%

The observation of neutrinoless double beta decay would establish lepton number violation and confirm the Majorana nature of neutrinos. In the standard light-neutrino exchange scenario, the decay rate is governed by the effective Majorana mass $\m$. For inverted ordering, this quantity is bounded from below at a level already within the reach of current and next-generation experiments. For normal ordering, it can be strongly suppressed, motivating experimental sensitivity beyond the IO region.

Using global-fit neutrino oscillation data which includes the first results from JUNO,
leading to an increased precision in the solar oscillation parameters,
we have updated the results of Ref.~\cite{Penedo:2018kpc}, i.e.~the conditions under which $\mno$ exceeds the millielectronvolt reference value. A sufficiently large or sufficiently small value of the lightest neutrino mass, $\mmin > 10.0$ meV or $\mmin < 0.2$ meV (for $3\sigma$ variations of parameters), is enough to guarantee $\mno > 1$ meV independently of the values of the unknown CP-violating phases. To surpass a higher threshold of $\mno > 5$ meV requires instead $\mmin > 20.8$ meV (for $3\sigma$ variations), 
corresponding to $\sum_i m_i > 0.097$ eV and
approaching the quasi-degenerate limit.

We have also considered scenarios in which the leptonic CP phases take specific values motivated by CP conservation or by generalized CP and flavour symmetries. In these cases, CP-violating phase choices that are still compatible with generalized CP symmetries now strictly imply $\mno > 1$ meV. Overall, our results reinforce the importance of extending neutrinoless double beta decay searches beyond the IO frontier. Should current and next-generation experiments yield null results, the continued exploration toward meV sensitivities remains well-motivated and essential for probing long-anticipated physics beyond the Standard Model.

%
\section*{Acknowledgements}
%

We would like to thank F.~Capozzi, E.~Lisi and A.~Marrone
for kindly sharing their data files with one-dimensional $\chi^2$ projections.
The work of S.T.P.~was supported in part by the European Union's Horizon Europe research and innovation programme under the Marie Sk\l{}odowska-Curie grant agreement No. 101086085-ASYMMETRY, by the Italian INFN program on Theoretical Astroparticle Physics and by the World Premier
International Research Center Initiative (WPI Initiative, MEXT), Japan.

\vfill
\clearpage
\footnotesize
\bibliographystyle{JHEPwithnote.bst}
\bibliography{bibliography}

\end{document}